\documentclass[pdflatex,sn-nature]{sn-jnl}

\usepackage{indentfirst}
\usepackage{caption}
\usepackage{setspace}
\usepackage{amsmath, amssymb, graphicx}
\usepackage{float}
\usepackage{pdfpages}
\usepackage{xcolor}
\usepackage{multirow}%
\usepackage{amsmath,amssymb,amsfonts}%
\usepackage{amsthm}%
\usepackage{mathrsfs}%
\usepackage[title]{appendix}%
\usepackage{xcolor}%
\usepackage{textcomp}%
\usepackage{manyfoot}%
\usepackage{booktabs}%
\usepackage{algorithm}%
\usepackage{algorithmicx}%
\usepackage{algpseudocode}%
\usepackage{listings}%

\usepackage{titlesec}

\titleformat{\title}{\normalfont\LARGE\bfseries}{\thesection}{1em}{}

\titleformat{\section}{\normalfont\Large\bfseries}{\thesection}{1em}{}
\titleformat{\subsection}{\normalfont\large\bfseries}{\thesection}{1em}{}

\begin{document}

\begin{center}
    \textbf{\LARGE Zero-shot Autonomous Microscopy for Scalable and Intelligent Characterization of 2D Materials}\\
    ~\\
    Jingyun Yang$^{1,\dagger}$, Ruoyan Avery Yin$^{2,\dagger}$, Chi Jiang$^1$, Yuepeng Hu$^1$, Xiaokai Zhu$^1$, Xingjian Hu$^1$, Sutharsika Kumar$^1$, Xiao Wang$^3$,  Xiaohua Zhai$^3$, Keran Rong$^3$, Yunyue Zhu$^4$, Tianyi Zhang$^4$, Zongyou Yin$^5$, Jing Kong$^4$, Neil Zhenqiang Gong$^1$, Zhichu Ren$^{6,*}$, Haozhe Wang$^{1,*}$\\
        \small 1 Department of Electrical and Computer Engineering, Duke University\\
    2 School of Computing, National University of Singapore\\
    3 Google Deepmind\\
    4 Department of Electrical Engineering and Computer Science, Massachusetts Institute of Technology \\
    5 School of Chemistry, Australian National University \\
    6 Department of Materials Science and Engineering, Massachusetts Institute of Technology\\
    ~\\
    $\dagger$ These authors contributed equally to this work.\\
    * Correspondence to: zc\_ren@alum.mit.edu, haozhe.wang@duke.edu
\end{center}

\section*{Abstract}
Characterization of atomic-scale materials traditionally requires human experts with months to years of specialized training. Even for trained human operators, accurate and reliable characterization remains challenging when examining newly discovered materials such as two-dimensional (2D) structures. This bottleneck drives demand for fully autonomous experimentation systems capable of comprehending research objectives without requiring large training datasets. In this work, we present ATOMIC (Autonomous Technology for Optical Microscopy \& Intelligent Characterization), an end-to-end framework that integrates foundation models to enable fully autonomous, zero-shot characterization of 2D materials. Our system integrates the vision foundation model (\textit{i.e.}, Segment Anything Model), large language models (\textit{i.e.}, ChatGPT), unsupervised clustering, and topological analysis to automate microscope control, sample scanning, image segmentation, and intelligent analysis through prompt engineering, eliminating the need for additional training. When analyzing typical MoS$_2$ samples, our approach achieves 99.7\% segmentation accuracy for single layer identification, which is equivalent to that of human experts. In addition, the integrated model is able to detect grain boundary slits that are challenging to identify with human eyes. Furthermore, the system retains robust accuracy despite variable conditions including defocus, color temperature fluctuations, and exposure variations. It is applicable to a broad spectrum of common 2D materials—including graphene, MoS$_2$, WSe$_2$, SnSe—regardless of whether they were fabricated via top-down or bottom-up methods. This work represents the implementation of foundation models to achieve autonomous analysis, establishing a scalable and data-efficient characterization paradigm that fundamentally transforms the approach to nanoscale materials research.

\section*{Main}
 Accurate and efficient characterization is critical for the research and applications of materials at the atomic scale. Advanced characterization tools, including microscopy, have made significant contributions to exploring new 2D materials and gaining fundamental understanding \cite{li2020general, ko2023operando, li2018vapour, najmaei2013vapour}. However, current microscopy methods involve manual inspection and interpretation, creating bottlenecks in research workflows and thus limiting the pace of nanomaterials research. Recent adoption of supervised learning algorithms for materials research, including analyzing microscopic images, has demonstrated high accuracy \cite{raccuglia2016machine, zhao2023learning, Xie2018, zeni2025generative, granda2018controlling, Kandel2023, schutt2017quantum, lu2018accelerated, Zhu2022, Masubuchi2019}, but these algorithms have a high demand for a large amount of high-quality training data, which is infeasible in emergent research fields, particularly for newly discovered atomic materials.

Foundation Models (FMs) including large language models (LLMs) \cite{openai2023gpt4} and vision foundation models (VFMs) have recently emerged as a transformative class of AI models, characterized by large-scale pre-training on diverse datasets and remarkable capabilities in natural language processing, knowledge reasoning, image analysis, and task generalization \cite{radford2021learning, bommasani2021opportunities, hou2024assessing, wang2023chatcad, fu2024drive, boiko2023emergent}. Due to their extensive pre-training on diverse datasets, FMs demonstrate zero-shot (no training) or few-shot (small dataset training) capabilities in various tasks\cite{kojima2022large}. This feature is particularly valuable for analyzing scientific results, where tasks often involve scarce experimental data and high annotation costs that make traditional supervised learning methods impractical.

In this work, we developed an integrated framework, ATOMIC, that integrates FMs to enable fully autonomous optical microscopy (OM) for zero-shot 2D materials characterization. Our system integrates the VFM (\textit{i.e.}, Segment Anything Model, SAM) with LLMs (\textit{i.e.}, ChatGPT) to create a comprehensive solution that encompasses microscope control, sample scanning, image segmentation, and intelligent analysis. Unlike conventional ML approaches, our method requires only prompts, but no labeled training data, while offering universal analysis across diverse materials with seamless integration into various experimental systems.

\subsection*{Large Model Controlled Microscopy}
ATOMIC processes natural language to generate microscopic analysis, as illustrated in Fig. 1a. The system comprises three synergistic core modules: Physical Control, which autonomously adjusts settings for high-quality image acquisition (Fig. 1b); Spatial Segmentation, which employs zero-shot SAM to generate masks representing potential material domains (Fig. 1c); and LLM-Supervised Clustering, which classifies segmentation masks using RGB channels to distinguish between monolayer, multilayer, and impurity regions (Fig. 1d). ATOMIC demonstrates high accuracy on CVD-grown MoS$_2$ flake size with a 0.026 Jensen-Shannon divergence against expert annotations (Fig. 1e). The model achieves 99.7\% pixel-wise accuracy and 99.9\% pixel-wise recall for monolayer MoS$_2$ (Fig. 1f), confirming its capability for autonomous materials characterization.

The Physical Control Module utilizes GPT to autonomously manage microscope operations, dynamically adjusting focus, exposure, and stage movement in real time to optimize imaging conditions. ZEISS Micro Toolbox (MTB) serves through a Flask-based API to GPT, facilitating seamless communication between the microscope and ATOMIC, thereby enabling precise and adaptive control.

Within this framework, a GPT-based agent (GPT-4o-mini, July 2024 release) interprets user-defined task instructions, decomposes them into actionable subtasks, and issues corresponding control commands (Fig. S1). This interaction allows for real-time decision-making, ensuring that the microscope adjusts imaging conditions without human intervention. By leveraging MTB’s hardware control capabilities and the GPT agent’s high-level reasoning, the system adaptively optimizes imaging parameters based on real-time feedback. A structured system prompt provides contextual awareness, defining the research domain and the GPT agent’s role in operating the microscope. Additionally, structured task prompts specify scanning paths and region of interest to enable inch-scale sample inspection. For instance, a typical field of view is captured in approximately 6 seconds, allowing a $1 \text{ cm} \times 1 \text{ cm}$ substrate to be scanned in ~6 minutes under 10× magnification. 


\subsection*{Spatial Segmentation}

Following GPT-controlled OM image acquisition, the Spatial Segmentation Module applies SAM for initial region segmentation. Specifically, SAM operates in mask generation mode, configured with thresholds for Intersection over Union (IoU), stability, and mask refinement, producing a set of candidate masks across the entire image without training. With its high precision in detecting material boundaries, SAM effectively distinguishes monolayer MoS$_2$ from substrate regions, even when edge contrast is exceptionally low. 

Fig. 2a–d compare conventional Canny edge detection with SAM's segmentation output. Unlike Canny detection, which struggles with continuous boundaries and primarily captures high-contrast multilayer regions, SAM successfully identifies all relevant flakes, including monolayers with low-contrast edges, and even intricate fractal-like structures challenging to human eyes. Moreover, SAM avoids issues including impurity misclassification, missing multilayer, and missing small flakes, encountered in conventional computer vision methods (see more discussion in Supplementary S3). 

SAM's zero-shot performance was evaluated by comparing the model's segmentation performance against human-labeled ground truth data. SAM generating initial candidate masks, followed by a filtering process to eliminate superfluous masks based on SAM-estimated IoU values. These IoU metrics were estimated by SAM's internal prediction head, which processes the feature correlation between the generated mask and the input image \cite{kirillov2023segment}. To validate SAM's zero-shot mask selection capability, we compared SAM-predicted IoU scores against human-labeled ground truth. Figure S2 demonstrates that SAM achieves 95 \% mean accuracy for selected masks, confirming the reliability of SAM's IoU-based selection mechanism. These results establish that zero-shot SAM effectively identifies 2D materials flakes without requiring domain-specific training. 

SAM exceeds human performance in specific tasks. Fig. 2e–h demonstrate SAM's capability to detect subtle grain boundary slits that are imperceptible to human eyes. While human require contrast enhancement and zooming in to discern some of these internal structures, SAM directly detects them from the full original image (Fig. 2g). These slits indicate areas where growth originated from separate nucleation sites that later merged, suggesting potential differences in crystal orientation or other underlying structural flaws. This ability to distinguish grain boundary slits is crucial in manufacturing 2D material devices, as these slits significantly deteriorate optical and electrical properties. Consequently, avoiding the incorporation of boundary-containing regions during device fabrication is essential for increasing the success rate \cite{cheng2015kinetic, hsieh2017effect}.

We developed a topological analysis approach that enhances SAM's ability to differentiate overlapping regions, thereby further increasing its segmentation accuracy. Although SAM generates highly precise segmentation, it often produces multiple overlapping masks when analyzing structures with complex topologies (Fig. 2i). In such cases, a single mask may encompass distinct regions requiring separation. To resolve this problem, we implemented a topological correction to eliminate mask overlaps and ensure discrete segmentation of each region. Fig. 2i illustrates a typical case where a vibrant blue multilayer MoS$_2$ region has grown within a monolayer structure. SAM produces two masks: one exclusive to the multilayer region and another covering both monolayer and multilayer regions (denoted red). As demonstrated in Fig. S14, we applied a topological subtraction to remove the multilayer mask from the combined mask, yielding a refined boundary (denoted green). Fig. 2j shows a significant decrease in standard deviation within the mask after correction, alongside a notable shift in mean RGB values, confirming improved isolation of a single 2D material region. As detailed in Supplementary S4, this topological correction procedure increased the classification accuracy from 77.92\% to 100\% for typical images, effectively eliminating errors caused by mask overlap and RGB miscalculations.

\subsection*{LLM-supervised Clustering}

To understand the material characteristics from SAM generated masks, we firstly perform color analysis for layer number identification. Masks are transformed into vectors in RGB (Red, Green, Blue) space, where vector values are represented by mean RGB values in each mask (Fig. 3a). RGB space was selected for its strong alignment with human visual perception, making it interpretable for large models. In addition, we applied similar analysis in HSV color spaces for comparison (see discussion in Supplementary S5). Analysis of 2D projections reveals that the Red channel plays a key role in distinguishing categories, showing clear separation in Red-Blue and Red-Green planes, while the Green-Blue plane exhibits more overlap, particularly between purple and orange categories. Consistently, R-channel noise primarily affects substrate-monolayer differentiation, whereas B-channel noise influences the distinction between monolayer and multilayer regions (see more discussion in Supplementary S6). This pattern aligns with the reflectance properties of few-layer MoS$_2$, where variations are most pronounced in the red-wavelength range due to excitonic absorption features \cite{li2018layer}. Based on the preceding RGB analysis, it becomes feasible to classify the SAM-generated masks into distinct labeled groups that correspond to different layer numbers.

We discovered that LLMs effectively supervise unsupervised algorithms, such as k-means clustering, enabling a fully autonomous process. Conventionally, the optimal number of clusters k is determined by human selection or the Elbow method, which identifies the point of maximum curvature in the distortion curve as illustrated in Fig. 3b (green line). However, the Elbow method often lacks precision, particularly when sample distributions are imbalanced and skew the distortion curve, or when impurities exhibit dramatically different colors from primary material classes. To address these issues, we developed a LLM-supervised k-means clustering approach that analyzes the distortion data and selects the optimal k value, as indicated by the red line in Fig. 3b. The LLM (GPT-4o, OpenAI, July 2024) was provided with the clustering distortion values and percentage change as input, with the objective of recommending an optimal k based on prompts. 

Our GPT-assisted k-means clustering demonstrates superior stability and accuracy in classification compared to the Elbow method, yielding results that more closely aligned with human-annotated ground truth. As shown in Fig. 3c, we compared k value selections over 35 images by human, GPT and Elbow algorithm with 15 trials for each analysis. Specifically, compared to the human-annotated ground truth, the GPT-4o model achieves same k selection in 29 out of 35 sample images (82.9\%), while the conventional Elbow method achieved a same selection in 11 out of 35 sample images (31.4\%). We found that Elbow algorithm shows mean absolute error of 0.83 comparing with human selection, while GPT-4o shows 0.17, as demonstrated in Fig. 3c error margin. In addition, we evaluate the k selection capability for different versions of GPT. GPT-4o outperforms GPT-o1-mini, exhibiting higher k selection accuracy (82.9\% vs. 28.6\%) and lower variance (0.783 vs. 1.025) across 15 independent queries for each of the 35 images, as illustrated by the error bars in Fig. 3c.

Fig. 3d-g provides a typical example wherein the Elbow method selects k = 3 while GPT refines this to k = 4. This critical adjustment allows GPT to successfully distinguish between impurities and multilayer MoS$_2$, while simultaneously correcting misclassifications of multilayer regions previously labeled as monolayer MoS$_2$ (Fig. 3e-f). In Fig. 3g, the difference map illustrates these corrections with blue regions indicating areas corrected by GPT, orange regions representing areas that were misclassified by both methods, and gray regions showing areas that were correctly classified by both approaches. The classification accuracy, calculated as the proportion of correctly classified masks before and after GPT-assisted k selection, improved from 67.80\% (40 out of 59 masks) to 98.30\% (58 out of 59 masks) in the example image. We demonstrate that GPT can serve as an agent for selecting optimal k values in conventional k-means clustering (see more discussion in Supplementary S7), comparable to human annotation. This capability transforms unsupervised learning into a fully autonomous process for image analysis within our ATOMIC model.

\subsection*{Robustness}

To evaluate the robustness of the image analysis enabled by SAM and GPT-supervised clustering, we tested its performance under defective microscopic imaging conditions (Fig. 4a). The input images were acquired under suboptimal conditions typically encountered during routine data collection, including color desaturation (reduction in color densities), defocus, white balance variations across different color temperatures (CTs), and exposure inconsistencies (both under- and overexposure). Accuracy values are calculated at both mask and pixel levels: mask-level accuracy represents the proportion of correctly classified masks, while pixel-level accuracy is weighted by the area coverage of each mask. As the analysis focuses exclusively on non-substrate categories, true negatives were excluded from both calculations. The original image achieved 95.7\% mask-level accuracy, while defective images maintained comparable performance as shown in Fig. 4b. Across all conditions, pixel-level accuracy remains above 99.5\%, demonstrating the system’s resilience to imaging variations (Fig. 4b). The layer number classification is further validated by the Atomic Force Microscopy images in Fig. S12.

Our system demonstrates consistent capabilities for analyzing diverse 2D materials. For mechanically exfoliated MoS$_2$, the system successfully identifies both layer numbers and distinguishes challenging features such as adhesive residuals (Fig. 4c Column 1 Row 2) and folded regions in monolayer flakes—structures (Fig. 4c Column 2 Row 3) typically difficult for human operators to consistently differentiate. For physical vapor deposition (PVD) grown SnSe, it distinguishes among few-layer, multilayer, and vertically grown samples. For chemical vapor deposition (CVD) grown WSe$_2$, it accurately identified monolayer, bilayer, and multilayer regions. For CVD-grown graphene, the system precisely identified monolayer, bilayer, and trilayer regions, as well as detected in-plane wrinkles (boundary between two bilayer masks as shown in Fig. 4c Column 6 Row 2). The results exhibit ATOMIC's universal application to analyzing various 2D materials with high accuracy. 

We compared our ATOMIC system with existing machine learning algorithms (Fig. 4d-e). Notably, our accuracy calculations include a specific portion of SAM generated masks — substrate regions initially misidentified by SAM as materials (false positives) but subsequently corrected through GPT-supervised clustering to align with ground truth. 
Specifically, as shown in Fig. 1c, when substrate areas are completely enclosed by boundaries, SAM categorizes these regions as positive masks, falsely indicating the presence of materials. However, through GPT-supervised clustering, these masks are correctly classified as substrate, yielding accurate results (Fig. 1d). This issue in SAM stems from its edge-based segmentation priorities, a limitation effectively remedied by our GPT-supervised clustering approach.
Our zero-shot approach achieves 99.7\%  pixel-level accuracy, with a 95\% confidence interval (CI) of 99.4\% - 100.0\%. This narrow confidence interval indicates consistent performance across all images, outperforming most conventional models despite requiring no training data. As shown in Fig. 4d and e, we summarize the accuracies and training dataset sizes for other machine learning models perviously reported to analyze 2D materials. Our ATOMIC framework achieves state-of-the-art performance with the highest accuracies at both mask and pixel levels, demonstrating a remarkable 99.7 \% pixel-level accuracy without requiring any training data, while maintaining its capability under challenging imaging conditions and across various 2D material species.

\section*{Discussion}
We develop a framework composed of foundation models to achieve autonomous microscopic characterization without training. The framework utilizes GPT to control the microscope, SAM to generate masks, and GPT-supervised k-means clustering to analyze 2D materials, demonstrating $>$99\% accuracy in understanding 2D materials. Comparing with conventional methods including machine learning algorithms, our large-model synergy offers key advantages:
(1) Elimination of reliance on extensive labeled training data, making it particularly valuable in emerging research fields with limited high-quality annotations.
(2) Fully autonomous operation, from microscope control to analysis output.
(3) Close-to-human high accuracy and detection of subtle features, such as grain boundary slits, that are challenging to human eyes.
(4) Robust performance under varying imaging conditions, fabrication methods, and diverse material species.

This work establishes a scalable, data-efficient framework for characterizing 2D materials, with implications extending throughout materials science and into adjacent scientific disciplines. The close-to-human high accuracy achieved in zero-shot conditions highlights the capability of foundation models to solve specific scientific problems without domain-specific training. In addition, the system's ability to detect defects beyond human eyes indicates the potential of minimizing human bias, thus advancing closed-loop materials discovery. The integration of foundation models and prompt engineering represents a promising avenue for advancing autonomous research in materials discoveries and beyond.

\section*{Methods}
\subsection*{Physical Controlling of the Microscopy}
The optical microscope used in this study is a Zeiss AxioScope 7, equipped with a Zeiss Axiocam 705 color camera (Carl Zeiss A.G., Oberkochen, Germany) and Zeiss ZEN 3.10 software. To enable autonomous control, we implemented a Python-based interface utilizing the \textbf{\texttt{pywin32}} package, which allows communication to the ZEN software via COM objects. 

We leverage GPT's function-calling capabilities to implement precise control and facilitate seamless translation between natural language prompts and microscope operations (Fig. S1). This mechanism allows the model to generate structured, executable commands rather than free-text outputs, ensuring accurate interpretation and execution of system functions based on user input.

We developed an approach that converts Google-style docstring-annotated Python functions into GPT-compatible JSON schemas. These schemas define the required parameters for valid API interactions, facilitating accurate interpretation by the GPT agent. By standardizing function descriptions, this approach enables dynamic invocation of complex control commands while minimizing manual intervention and potential errors. The implementation details are provided in Supplementary S1.

Additionally, white balance and contrast parameters are fine-tuned through the ZEN software to ensure consistency in imaging conditions. Captured images are stored in a predefined format with a structured naming format to facilitate subsequent analysis.

\subsection*{Spacial Segmentation by SAM}
We employ the SAM model using the \textbf{\texttt{sam\_vit\_h\_4b8939}} checkpoint to segment the acquired microscopy images. The model runs on an NVIDIA T4 GPU through Google Cloud Platform, leveraging its tensor cores to accelerate the segmentation process.

The input to SAM consists of raw microscopic images in RGB format. The model operates with a Vision Transformer (ViT-H) backbone, characterized by a hidden size of 1280, 16 attention heads per layer, and 32 transformer layers. The image encoder processes images at a resolution of 1024 × 1024, which is internally resized to match the model’s default configuration. The segmentation masks are generated using the key parameters detailed in Supplementary S9.1.

Once SAM generates multiple masks for an image, we apply an overlap-based filtering strategy to remove redundant segmentation as part of topological correction. Specifically, masks that significantly overlap with other remaining masks after subtracting the intersection are considered redundant and are filtered out. This step ensures that inclusion relationships (e.g., few-layer structures inside a monolayer region) are preserved while eliminating unnecessary duplicate segmentation (Fig. S14). Conversely, if the overlap falls below a certain threshold, no filtering is applied, as these cases are considered minor boundary ambiguities rather than redundant masks. The detailed implementation of this correction process is described in Supplementary Algorithm 1.

This segmentation process allows for precise mask generation while preserving hierarchical relationships among different MoS$_2$ layers, ensuring that monolayer, few-layer, and multilayer structures are accurately described.

\subsection*{Prompt Engineering}
To initiate and manage the overall microscope control process, we implement a structured system prompt that defines operational context, available commands, and execution protocols. This framework guides the front-end GPT agent through the scanning process using step-by-step instructions, explicit function definitions, and detailed task requirements. The exact system prompt is detailed in Supplementary S9.2. The prompt incorporates conditional logic for position selection, hardware coordination protocols, and clear completion criteria, collectively enhancing interpretability and execution efficiency.

The task execution prompt employs a structured format that enables the back-end GPT agent to autonomously search areas of interest. This prompt features explicit dimensional parameters, sequential action guidelines, contextual spatial anchoring, and conditional directives for optimal region selection. The exact prompt used is detailed in Supplementary S9.2. These elements ensure precise execution of complex microscopy tasks while maintaining spatial consistency.

For image classification, we utilize a specialized prompt incorporating the Elbow Method with a fixed initial cluster center based on the calibrated RGB color of the substrate. This prompt provides explicit classification criteria, quantitative decision thresholds, and rule-based constraints for determining optimal cluster counts. A structured version of this prompt is provided in Supplementary S9.2. This approach delivers precise, context-aware image segmentation results while maintaining analytical rigor.

\subsection*{Materials Fabrication}
Monolayer MoS$_2$ was synthesized in a 1-inch quartz tube within a single-zone furnace. Sulfur powder (99.98\%, Sigma Aldrich) was placed in an alumina crucible positioned upstream in the tube. A homogeneous mixture of MoO$_3$ (99.5\%, Sigma Aldrich) and NaCl (99.5\%, Sigma Aldrich) was filled in a separate alumina boat and positioned at the furnace center. A clean SiO$_2$/Si substrate was positioned face-down above the MoO$_3$ powder. The tube was initially purged with 500 sccm Ar to eliminate air. The system temperature was raised to 780 °C over 30 minutes using 120 sccm Ar as the carrier gas. Monolayer MoS$_2$ synthesis occurred at 780 °C for 10 minutes, after which the system was allowed to cool naturally to ambient temperature. 

Exfoliated MoS$_2$ flakes were mechanically exfoliated from commercially molybdenite crystals (470MOS2S-AB, SPI Supplies) using the Scotch Magic Tape, and transferred onto the Si substrates with 300 nm layer of SiO$_2$.

Tin selenide (SnSe) nanoflakes were synthesized onto mica substrates using the physical vapor deposition (PVD) method. SnSe powder (99.999\%, Thermo Scientific Chemicals) was used as the precursor, placed in a quartz boat at the center of a 1-inch furnace. The mica substrate was freshly cleaved to expose a clean surface, annealed at 400 °C in air for 10 min, and positioned downstream in the furnace approximately 10 cm from the precursor. The PVD system was evacuated to 12 mTorr using a mechanical pump, and the furnace was heated to a center temperature of 430 °C in 10 min with a carrier gas flow of 65 sccm argon (Ar) and 5 sccm hydrogen (H$_2$). The growth process was maintained at 430 °C for 60 min, after which the furnace was rapidly cooled to room temperature by opening the furnace cap and applying an external fan.

2D WSe$_2$ flakes were synthesized on SiO$_2$ (300 nm)/Si by a chemical vapor deposition (CVD) approach that was previously developed \cite{liu2024boosting}. First, an aqueous solution containing ammonium metatungstate hydrate (2 mg/mL) and sodium chloride (0.5 mg/mL) was spin-coated onto a piece of SiO$_2$/Si wafer. The wafer was then loaded in a tube furnace, and the furnace temperature was ramped to 850 °C and held for 5 min for the selenization process. 50 sccm of argon and 5 sccm of hydrogen were used as carrier gas during the selenization reaction. After the synthesis process, the WSe$_2$ sample was fast cooled under 200 sccm of argon.

Bottom-up graphene samples were fabricated using Cu foils (Alfa Aesar no. 46365) folded into pocket structures and loaded into a low-pressure chemical vapor deposition (LPCVD) system. The system temperature was gradually increased to 1,050°C over 30 min while maintaining chamber pressure at 50 mTorr. After temperature stabilization at 1,050°C for 30 min, 50 sccm of H$_2$ gas was introduced. One minute following hydrogen introduction, 1 sccm of CH$_4$ was added to initiate bilayer graphene (2LG) growth. These synthesis parameters were maintained for 1 hr to complete the growth process.

\section*{Acknowledgements}
J.Y. and H.W. acknowledge the GCP Credit Award from the Google PaliGemma Academic Program. J.Y., X.Z.(Zhu), X.H., H.W. acknowledge the start-up fund from Duke University. C.J. acknowledges the support of Sigma Xi Summer Award from Swarthmore College. Y.Z., T.Z. and J.K. acknowledge the support by the Semiconductor Research Corporation Center 7 in JUMP 2.0 (award no. 145105-21913).

\section*{Contributions}
H.W. and Z.R. conceived the idea and supervised the work. J.Y., R.A.Y., Z.R., and H.W. developed the microscope controlling module, J.Y., Y.H., S.K., N.Z.G., and H.W. developed the segmentation module. J.Y., C.J., and H.W. developed the GPT-supervised clustering module. X.Z.(Zhu), X.H., Y.Z., T.Z., C.J., J.K., and H.W. fabricated the 2D materials used for characterization. X.W., X.Z.(Zhai), K.R., and Z.Y. offered advice during the research.  J.Y., R.A.Y., X.H., X.Z.(Zhu), Z.R. and H.W. drafted the manuscript. All authors reviewed and revised the manuscript.




\newpage
\begin{figure}[H]
    \centering
    \includegraphics[width=1\textwidth]{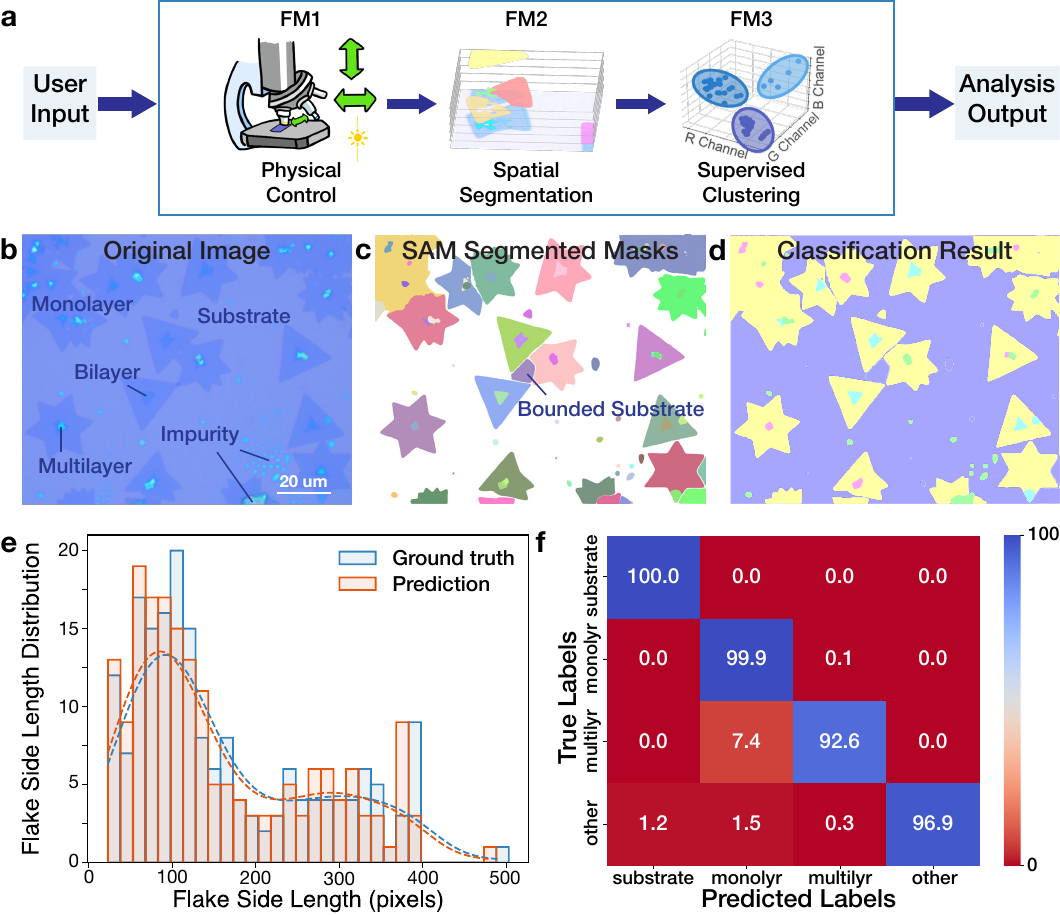}
    \caption{\textbf{ATOMIC framework to autonomously analyze 2D materials.} \textbf{a,} Schematic workflow of the ATOMIC framework, illustrating the synergistic integration of foundation models to enable autonomous microscopy. The system combines GPT for hardware control, Segment Anything Model (SAM) to generate segmentation masks, and GPT-supervised clustering to identify materials species. This multi-model approach enables microscope control, decision-making, and autonomous analysis to achieve self-directed microscopic imaging and characterization without human intervention.
    \textbf{b,} Optical microscopic image of MoS$_2$ crystals synthesized via chemical vapor deposition, captured using our autonomous microscopy system. The image exhibits characteristic monolayer, bilayer, and multilayer regions with distinct optical contrast. \textbf{c,} Segmentation masks generated by the Segment Anything Model (SAM) highlighting potential material regions, overlaid on the original imgage. \textbf{d,} Material classification results obtained through GPT-supervised clustering, distinguishing between substrate, monolayer, bilayer, multilayer regions, and impurities.
    \textbf{e,} Size distribution histogram of triangular monolayer MoS$_2$ flakes, showing side length measurements. The dashed line represents the kernel density estimate (KDE) of the distribution.
    \textbf{f,} Pixel-level confusion matrix for the automated identification of CVD-grown MoS$_2$ crystal regions. Labels: "monolyr" = Monolayer, "multilyr" = Multilayer, "substrate" = Substrate, "other" = Impurities or unrelated regions. }
    \label{fig1}
\end{figure}

\begin{figure}[H]
    \centering
    \includegraphics[width=1\textwidth]{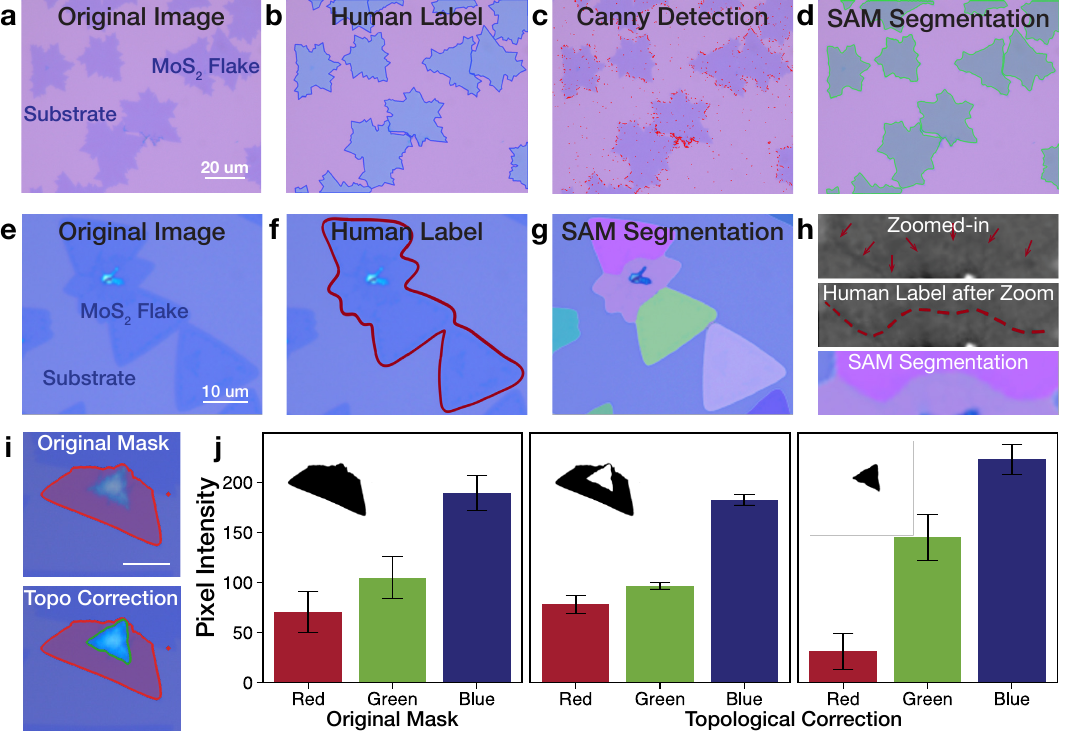}
    \caption{\textbf{Spatial segmentation enabled by Segment Anything Model and Topological Correction.} \textbf{a,} Original optical microscope image showing MoS$_2$ flakes on a substrate. \textbf{b-c,} Human annotated labels and Canny edge detection results. \textbf{d,} SAM Segmentation results with clear, continuous and detailed boundaries. \textbf{e,} Original optical micrograph showing MoS$_2$ flakes on substrate, \textbf{f,} Human annotation identifying a single continuous flake, and \textbf{g,} SAM-generated segmentation revealing four distinct flakes separated by grain boundary slits. Image magnified 2.5× from original acquisition at 100× magnification.
    \textbf{h,} Image comparison revealing a grain boundary slit that is directly detected by SAM at low magnification but only visually observable by human after image enhancement (zoom-in, increase contrast, and transfer to gray scale).
    \textbf{i,} Segmented masks before (top) and after (bottom) topological correction, with green line highlighting the corrected multilayer region boundary. Topological correction effectively compensate for SAM's limit in spatial capability. Scale bar=10$\mu$m.
    \textbf{j,} Mean RGB value with standard deviation of segmentation masks before (left) and after (middle and right) topological correction. Significant shift in mean RGB value and reduction in standard deviation shows enhanced precision in segmentation. }
    \label{fig2}
\end{figure}

\begin{figure}[H]
    \centering
    \includegraphics[width=1\textwidth]{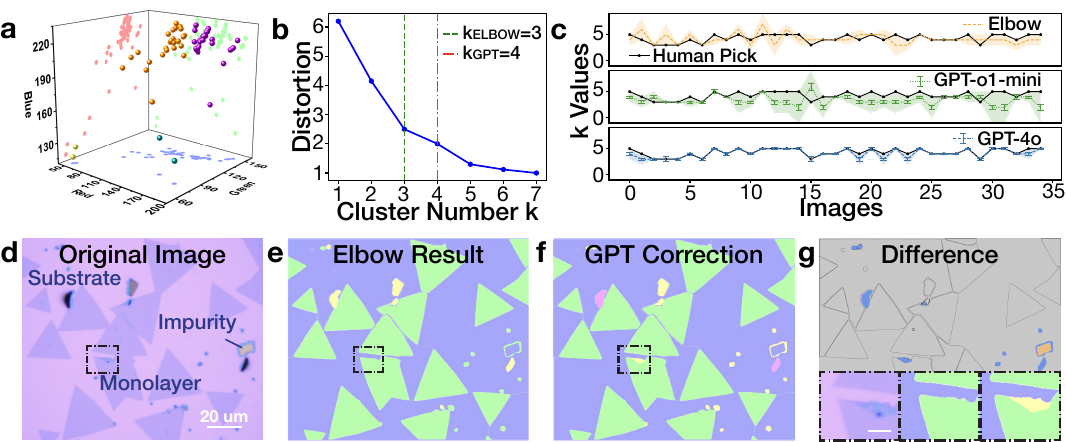}
    \caption{\textbf{GPT-supervised k-means clustering.} \textbf{a,} Three-dimensional visualization of mean RGB values for each segmented region plotted in RGB color space, with orthogonal projections displayed on the RG, GB, and RB planes to understand the contribution of each RGB channel to materials classification.
    \textbf{b,} Distortion analysis of a representative microscopic image showing the relationship between distortion and cluster number $k$. The optimal $k$-values identified by the conventional Elbow method (k=3, green) and GPT-assisted analysis (k=4, red) are indicated.
    \textbf{c,} Comparison of optimal cluster number $k$ selection across multiple sample images using three methods: Elbow method (orange, top figure), GPT-o1-mini (green, middle figure), and GPT-4o (blue, bottom figure), with human-selected values (solid black line in each figure) as baseline. The x-axis represents different image indices, while the y-axis shows selected $k$ values. Shaded regions indicate twice the margin of error from human-selected values for visualization purposes. Error bars represent standard deviations across 15 repeated queries for each image using GPT models, with modal values plotted as data points. GPT-4o demonstrates robust alignment with human expert decision in $k$ selection, exhibiting highly consistency and stability across multiple test samples. \textbf{d,} Original micrograph showing monolayers, impurities, and substrate, with a dashed frame highlighting a multilayer region identifiable by human experts. \textbf{e,} Material classification results using Elbow method-selected $k$ value, showing failure to detect the multilayer region marked in black frame. \textbf{f,} Classification results using GPT-selected k value, successfully detecting the multilayer region within the framed area, demonstrating superior alignment with human expert identification. 
    \textbf{g,} Spatial difference map highlighting classification discrepancies between Elbow method and GPT-assisted clustering approaches. Inset panels show magnified views of the framed region in panels \textbf{d-f}: Original microscopic image (left); Elbow method classification results incorrectly identifying multilayer regions as monolayer (middle); GPT-supervised classification results accurately distinguishing multilayer from monolayer regions (right). Scale bar=5$\mu$m. 
    }
    \label{fig3}
\end{figure}

\begin{figure}[H]
    \centering
    \includegraphics[width=1\textwidth]{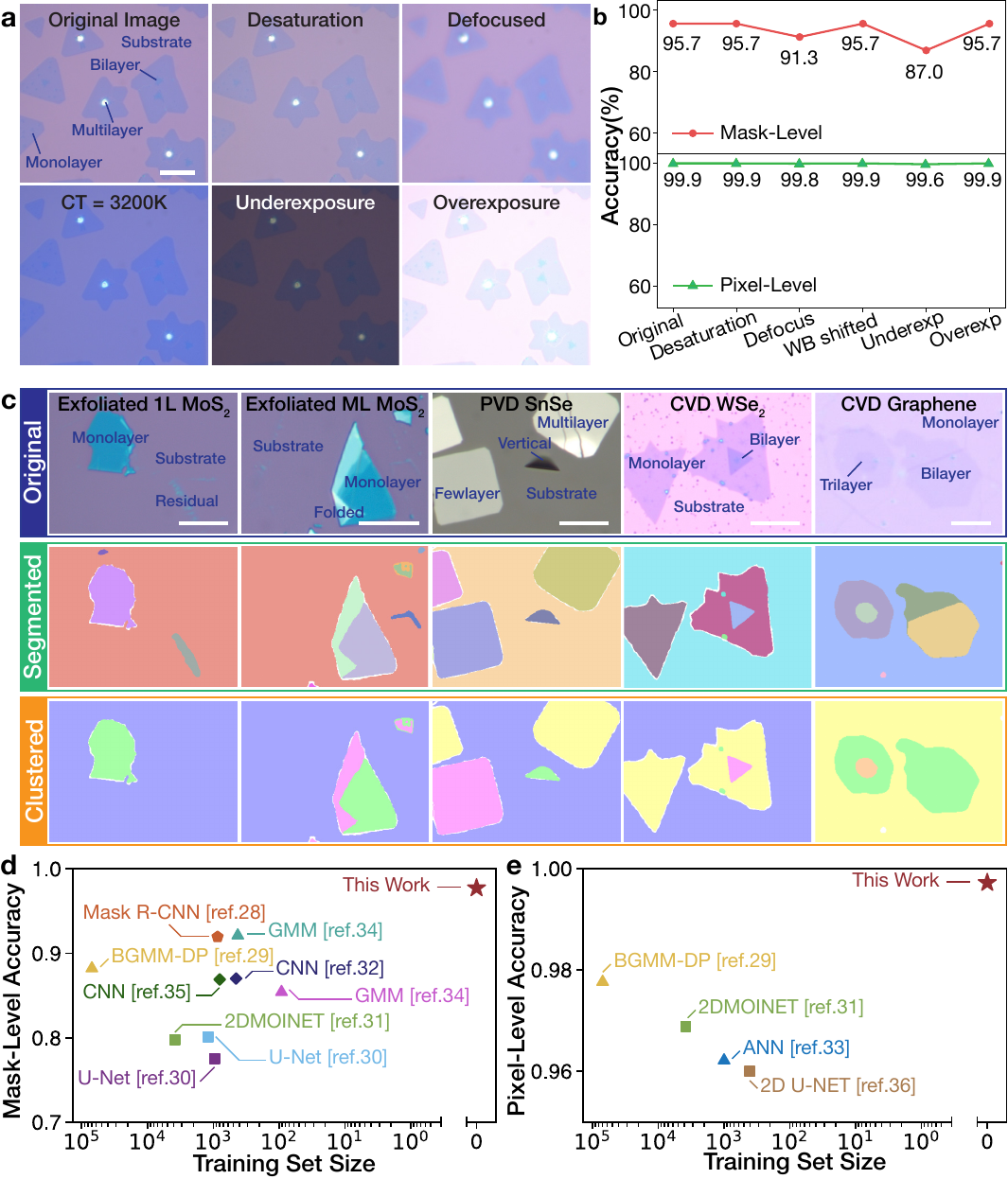}
    \caption{\textbf{Robustness of the foundation model synergy for 2D materials analysis.} \textbf{a,} Optical images of MoS$_2$ flakes acquired under varying imaging conditions: standard acquisition (top left), color desaturation (20\% intensity reduction, top middle), defocus (z-axis shift, top right), white balance shift (color temperature changed from 5500K to 3200K, bottom left), underexposure (24\% of standard exposure time, bottom middle), and overexposure (295\% of standard exposure time, bottom right). Layer thickness at the center of the standard captured image (top left) was validated by Atomic Force Microscopy (Fig. S13). Scale bar=10$\mu$m. }
    \label{fig4}
\end{figure}

\begin{figure}[H]
    \centering
    \caption*{\textbf{b,} Classification accuracy under each imaging condition assessed at both mask level (top panel) and pixel level (bottom panel). 
    \textbf{c,} Autonomous analysis of diverse 2D materials: mechanically exfoliated monolayer MoS$_2$ (column 1), exfoliated multilayer MoS$_2$ (column 2), PVD-grown SnSe (column 3), CVD-grown WSe$_2$ (column 4), and CVD-grown graphene (column 5). Top row shows original micrographs with human annotations; middle row displays SAM segmentation with topological corrections; bottom row presents layer classification after GPT-supervised clustering. Color code: substrate (purple), mono- or thinnest layer (yellow), bilayer (green), multilayer (pink). Scale bar=10$\mu$m. 
    \textbf{d-e,} Comparison of ATOMIC system performance with machine learning (ML)-based approaches. \textbf{d,} Mask-level accuracy comparison between our zero-shot ATOMIC system and previous ML methods, plotted versus training dataset size (x-axis). Our approach achieves 97.6\% accuracy without requiring training data. Detailed analysis provided in Supplementary S2. \textbf{e,} Pixel-level accuracy comparison with previously reported ML methods, demonstrating ATOMIC's superior performance (99.7\%). All values are adopted directly from the literature, with model type indicated for each data point \cite{masubuchi2020deep, masubuchi2019classifying, saito2019deep, han2020deep, greplova2020fully, zhu2022artificial, uslu2024open, dong20213d, leger2024machine}. The stars represent our work. }
    \label{fig4}
\end{figure}

\end{document}


\begin{center}
    \textbf{\LARGE Supplementary Information}
    \textbf{\Large \\ Zero-shot Autonomous Microscopy for Scalable and Intelligent Characterization of 2D Materials} \\
\end{center}

\begin{center}
    Jingyun Yang$^{1,\dagger}$, Ruoyan Avery Yin$^{2,\dagger}$, Chi Jiang$^1$, Yuepeng Hu$^1$, Xiaokai Zhu$^1$, Xingjian Hu$^1$, Sutharsika Kumar$^1$, Xiao Wang$^3$,  Xiaohua Zhai$^3$, Keran Rong$^3$, Yunyue Zhu$^4$, Tianyi Zhang$^4$, Zongyou Yin$^5$, Jing Kong$^4$, Neil Zhenqiang Gong$^1$, Zhichu Ren$^{6,*}$, Haozhe Wang$^{1,*}$\\
        \small 1 Department of Electrical and Computer Engineering, Duke University\\
    2 School of Computing, National University of Singapore\\
    3 Google Deepmind\\
    4 Department of Electrical Engineering and Computer Science, Massachusetts Institute of Technology \\
    5 School of Chemistry, Australian National University \\
    6 Department of Materials Science and Engineering, Massachusetts Institute of Technology\\
    ~\\
    $\dagger$ These authors contributed equally to this work.\\
    * Correspondence to:  zc\_ren@alum.mit.edu, haozhe.wang@duke.edu
\end{center}

\renewcommand{\contentsname}{\large Table of Contents}
\tableofcontents

\newpage

\section{Physical Controlling of the Optical Microscope}

As shown in Fig. S1, in our autonomous optical microscope system, we employ the CallingGPT framework to convert docstring-annotated functions into GPT-compatible JSON, enabling a two-tier GPT agent architecture that translates natural language commands into precise microscope operations via standardized API interactions.
The system operates through a two-tier agent architecture that utilizes GPT's function-calling mechanism \cite{openai2023functioncalling} to enhance automation efficiency and the CallingGPT framework \cite{CallingGPT} to process and execute microscope control commands efficiently:

- The front-end GPT agent interprets natural language input and extracts high-level control commands. It then transmits these structured commands to the microscope system through a Flask-based web framework \cite{ren2023crest}. The structured format ensures compatibility with the downstream function-calling process, enabling the agent to handle diverse input types effectively.

- The back-end GPT agent receives these structured commands and utilizes the CallingGPT-generated JSON schemas to accurately map each command to its corresponding microscope function. Through GPT's function-calling capabilities, the agent dynamically selects and executes the relevant functions (e.g., auto-focus, auto-exposure, stage movement). This approach minimizes ambiguity and ensures that each instruction is accurately translated into executable API calls.

This hierarchical command-processing pipeline enables seamless translation of natural language instructions into precise instrument operations through standardized API interactions.

\section{Accuracy analysis of our foundation model synergy}
The model demonstrates high performance at both mask and pixel levels, with strong precision and recall across most classes. As evident in Fig. S3, Substrate classification reaches 100\% precision and recall. Monolayer achieves high precision (99.4\%) but slightly lower recall due to misclassified Multilayer instances (24 masks, 310,422 pixels). Multilayer exhibits high recall (93.5\%) but lower precision due to confusion with Monolayer. The Other class maintains $>$95\% precision and recall, with minor misclassification. The primary issue lies in distinguishing Multilayer from Monolayer, affecting both recall and precision. Improving feature differentiation, data augmentation, and weighted loss adjustment could enhance model performance. This analysis is based on a total of 28 images and 1337 masks, providing a comprehensive evaluation of model performance. 

For monolayer accuracy, at the mask level, direct aggregation of the 1337 masks yields an accuracy of 97.6\%, while averaging across the 28 images results in an accuracy of 96.2\% (95\% CI: [93.5\%, 98.9\%]). Pixel-level accuracy consistently achieves 99.7\%. The 95\% confidence interval for mask-level accuracy is relatively narrow. This indicates consistent performance across various images and highlights the robustness of our approach. Notably, the direct aggregation of 1337 masks results in a higher accuracy of 97.6\%, suggesting that a small number of high-accuracy images may positively skew the overall accuracy estimate. However, the average accuracy across images (96.2\%) provides a more reliable reflection of the model's capability.

The discrepancy between mask-level and pixel-level performance, quantified in Table S1 and S2,  arises due to differences in the spatial distribution and size of instances within each class. The Other class typically consists of smaller masks, meaning that while the majority of instances are correctly classified at the mask level, even minor misclassifications result in a disproportionately high number of incorrectly labeled pixels, leading to reduced pixel-wise performance. Conversely, the Multilayer class generally comprises larger masks, so any misclassification affects a greater number of pixels, amplifying its impact in pixel-wise metrics. As a result, the Other class exhibits higher accuracy at the mask level but lower performance at the pixel level, whereas the Multilayer class demonstrates the opposite trend.
\section{Comparison of SAM Segmentation vs. Traditional Computer Vision (CV) Methods}

Fig. S4a presents a typical optical microscope image of CVD-grown MoS$_2$ captured under a 100× objective. The image contains distinct regions corresponding to the substrate, monolayer, multilayer, and impurities, as well as artifacts caused by lens contamination. In direct clustering, contamination spots were incorrectly segmented as a separate class (Fig. S4b, upper red box), while actual impurities were missed (lower red box). Here, each pixel is assigned the RGB value of its cluster center, producing a visually distinct representation of the clustered regions. In contrast, when using SAM for initial mask generation before clustering, contamination spots were sometimes assigned masks but were later correctly reassigned to the substrate class, preventing impurity omission. This improvement likely arises because SAM-based classification is independent of mask size, ensuring that small but distinct features are not disregarded. 

When applying traditional contour detection to the direct clustering results, only the most prominent outer edges were detected, failing to identify multilayer regions (Fig. S4c, left red box) and omitting many small flakes (right red box). This contour detection is applied directly to the RGB-mapped clustering result from Fig. S4b, meaning that errors in clustering propagate into edge detection, further limiting its accuracy. Moreover, clustering alone often merges adjacent regions or misclassifies fine structures, whereas SAM ensures a more precise and adaptive segmentation by delineating boundaries before classification. Unlike clustering, which can blur original edges, SAM preserves all edge details, ensuring finer structural integrity. 

To achieve more accurate classification, instead of automatically selecting the number of clusters \( k \), we manually set \( k = 4 \). Ideally, these four categories should correspond to the substrate, monolayer, multilayer, and impurities. However, in this case, directly classifying each pixel resulted in the substrate being split into two separate categories while still failing to distinguish impurities. To facilitate visualization, we applied high-contrast coloring to the clustering results. As shown in Fig. S4d, a noticeable circular region appears in the center of the image, which is due to overexposure caused by the microscope’s illumination system. Although the microscope software attempts to compensate for these effects, subtle variations remain detectable during image analysis.  

In contrast, as demonstrated in Figs. S4e and S4f, SAM-based classification successfully delineated the desired boundaries, distinguishing between monolayers, multilayers, impurities, and parts of the substrate. This improvement ensures that in the subsequent clustering step, the masks are correctly classified into four categories. The red boxes highlight two key corrections: the upper red box shows the successful classification of lens contamination as part of the substrate, while the lower red box indicates the proper identification of impurities. These results highlight the necessity of SAM in achieving accurate and reliable feature segmentation, particularly for complex structures where conventional methods fail.

\section{Importance of Topological Correction on SAM Generated Masks}

Each mask in the clustering results is annotated with its mask ID and the mean RGB value of its constituent pixels. From the magnified region in Fig. S5b, it is evident that the blue channel value of the monolayer region is significantly higher than that of other monolayer areas, placing it between monolayer and few-layer classifications. This discrepancy arises because the mask assigned to this monolayer region also includes an internal few-layer area, leading to inaccurate RGB computation. Such overlaps are common in SAM-generated masks, as SAM is unable to segment masks with topological holes. However, in CVD-grown MoS$_2$, the growth of few-layer structures on top of monolayer regions is a frequent phenomenon. In some cases, SAM even produces two completely overlapping masks. Additionally, multiple instances where the substrate is misclassified as a monolayer (highlighted by the red dashed box) are observed. This misclassification results from inaccurate RGB computation, which skews the cluster centroids and affects segmentation accuracy.

To address this issue, we implemented a correction step that systematically traverses all masks and filters out overlapping regions. During this process, overlapping portions are removed from their original masks and stored as separate new masks to ensure that each mask contains only one distinct material category. This correction step increases the total number of masks but significantly enhances classification accuracy. Notably, this method not only corrects errors introduced by SAM’s inherent limitations but also ensures that multilayer growth patterns are more accurately captured in further clustering analysis. After applying this correction, the classification accuracy improved from 77.92\% to 100\% in this example image, successfully eliminating errors caused by mask overlap and RGB miscalculations (Fig. S5c). This correction method can be extended beyond MoS$_2$ analysis to other materials with complex growth structures, where precise segmentation is crucial for understanding layer distribution and material interactions.

\section{Comparison of RGB Color Space vs. HSV Color Space}
For a sufficiently complex original image shown in Fig. S6a, we conducted clustering analysis in both the RGB color space and the HSV-transformed space. The results in Fig. S6b-c indicate that for CVD-grown MoS$_2$ samples, HSV-based clustering can effectively distinguish light yellow impurities, though not with perfect accuracy. However, it fails to differentiate between monolayer and multilayer regions, a limitation that was consistently observed across multiple images.  

On the other hand, clustering in the RGB space successfully separates monolayer regions from the internal multilayer structures. While some confusion arises between light yellow impurities and the substrate, this does not hinder the analysis of monolayer coverage and multilayer growth patterns. Thus, despite the trade-offs, RGB-based clustering proves more suitable for studying the distribution and morphology of monolayer and multilayer MoS$_2$ in optical microscopy images.

\section{Analysis of RGB Channel Importance}
Fig. S7a-b displays the original microscope image, and the clustering result, where the purple-blue regions represent the substrate, the red regions indicate monolayer area, and the green regions correspond to multilayer parts. Subsequently, Gaussian noises are added to each of the RGB channels of the image at a specified ratio to analyze the role and importance of each channel in the clustering process.

As shown in Fig. S8, although the perturbation in the 3 \% R-channel is barely perceptible to the naked eye, it still introduces noticeable confusion between the substrate and monolayer regions, as evidenced by the confusion matrix. In contrast, the G-channel perturbation is the most visually apparent, yet it has a relatively minor impact on the clustering results. Interestingly, the B-channel perturbation introduces subtle changes, slightly affecting the boundary distinction between monolayer and multilayer regions. These observations highlight the varying degrees of sensitivity each channel contributes to the clustering process.

As shown in Fig. S9, the 5 \% perturbation in the R-channel starts to introduce noticeable confusion, which becomes apparent to the naked eye. In Fig. S9d, many substrate points are misclassified as monolayer, and vice versa. This confusion is further confirmed by Fig. S9g, where the 5 \% R-channel noise causes a significant overlap between substrate and monolayer regions. In contrast, the 5 \% G-channel perturbation continues to have minimal impact on the clustering results, maintaining overall classification stability.

However, the 5 \% B-channel perturbation introduces pronounced confusion between monolayer and multilayer regions. As observed in Fig. S9f and S9i, a large number of monolayer points are misclassified as multilayer. This is likely due to the close similarity in their B-channel values—multilayer regions typically show an RGB of (23, 207, 252), while monolayer regions have (109, 150, 223). The similarity in the B-values makes these layers more susceptible to misclassification under noise. Additionally, B-channel noise appears to distort the boundary definition, further complicating the clustering.

These results underscore the varying sensitivity of each RGB channel in the clustering process, with the B-channel playing a critical role in distinguishing multilayer structures.

As shown in Fig. S10, the impact of 7 \% noise perturbation becomes more pronounced. The 7 \% R-channel noise introduces clearer confusion between the substrate and monolayer regions. In Fig. S10d, a significant number of substrate points are incorrectly classified as monolayer, while some monolayer points are misclassified as substrate. This confusion is further evidenced by Fig. S10g, where the misclassification pattern intensifies compared to noise-free results. Additionally, at this noise level, confusion between monolayer and multilayer regions also emerges, indicating that 7 \% R-channel noise begins to affect layer distinction beyond just the substrate.

The G-channel perturbation, while relatively less impactful, starts to introduce a degree of confusion between monolayer and multilayer regions when noise level increase to 7 \%, as seen in Fig. S10e and S10h. Although the overall clustering remains largely consistent, some monolayer points are misclassified as multilayer, suggesting that G-channel information contributes modestly to layer differentiation at higher noise levels.

In contrast, the 7 \% B-channel noise leads to severe confusion between monolayer and multilayer regions. As observed in Fig. S10f and S10i, the clustering result shows that monolayer and multilayer regions become almost indistinguishable, with the noise completely obscuring the boundary between these two layers.

These findings indicate that R-channel noise primarily affects the distinction between substrate and monolayer, while B-channel noise significantly impacts the differentiation between monolayer and multilayer regions. G-channel noise introduces minimal confusion but still contributes slightly to misclassification under higher noise levels.

\section{Classification Comparison between Elbow Method and GPT supervision}
In the main text, we demonstrated the issue of the Elbow method selecting overly small k values, leading to imprecise classification. Here, we present additional results showing that selecting excessively large k values can also result in inaccurate classification. As shown in Fig. S11a-c, when k=5, the multilayer region is further subdivided into two classes, which is unnecessary from a human perspective. Additionally, the substrate is incorrectly split into two classes, which hinders precise analysis of the growth outcome. In contrast, k=3 effectively distinguishes the multilayer, monolayer, and substrate, providing a clearer and more meaningful segmentation that aligns well with human judgment. This observation highlights the tendency of the Elbow method to overfit when the k value is set too high, compromising the interpretability of the results.

Fig. S11d-f illustrate a scenario where both the Elbow method and GPT selections agree with the human annotation. This typically occurs when the images are simpler, containing well-separated and distinguishable regions. Under such conditions, both methods are capable of producing accurate classifications. However, it is important to note that the agreement between the two methods is not a guarantee of correctness in all cases. The simpler image structure reduces the complexity of segmentation, but the Elbow method may still struggle with more challenging or noisy data where GPT's adaptability provides a distinct advantage.

\section{AFM Characterization of Thickness}
The AFM images in Fig. S12-13 illustrate the morphology and thickness characterization of monolayer, bilayer, and multilayer structures. These images validate the classification results of the ATOMIC synergy. The distinct height profiles confirm the model's ability to accurately differentiate between monolayer, bilayer, and multilayer structures, demonstrating its effectiveness in layer classification.

\section{Supplementary Methods}

\subsection{Spacial Segmentation by SAM and topological analysis}
\textbf{Key parameters for SAM:}

\begin{itemize}
    \item \textbf{\texttt{points\_per\_side} = 32}: Defines the number of grid points sampled per side for mask proposal generation.
    \item \textbf{\texttt{pred\_iou\_thresh} = 0.88}: Sets the IoU prediction threshold; masks with an IoU confidence score below this threshold are discarded.
    \item \textbf{\texttt{stability\_score\_thresh} = 0.95}: Filters out masks based on stability score to retain only robust segmentations.
    \item \textbf{\texttt{box\_nms\_thresh} = 0.7}: Ensures non-maximum suppression (NMS) is applied to overlapping bounding boxes with IoU greater than 0.7.
    \item \textbf{\texttt{crop\_n\_layers} = 1}, \textbf{\texttt{crop\_nms\_thresh} = 0.7}: Parameters controlling mask cropping behavior, which refine predictions in overlapping regions.
    \item \textbf{\texttt{min\_mask\_region\_area} = 0}: Prevents filtering out small masks, ensuring all detected structures are retained.
\end{itemize}

\textbf{Algorithmic structure of topological correction:} 
\begin{algorithm}[H]
    \caption{Topological Correction}
    \begin{algorithmic}[1]
    \State \textbf{Input:} Binary masks \texttt{masks}, min overlap \texttt{min\_percentage}
    \State \textbf{Output:} Filtered masks
            
    \State Initialize \texttt{filtered\_masks} $\gets$ \texttt{masks}, \texttt{updated} $\gets$ \textbf{True}
            
    \While{\texttt{updated}}
        \State \texttt{updated} $\gets$ \textbf{False}, \texttt{new\_masks} $\gets$ [], \texttt{masks\_to\_remove} $\gets$ \{\}               
    \For{$i = 1$ to $N$}
            \If{\texttt{filtered\_masks}[i] is None} \textbf{continue} \EndIf
            \For{$j = i+1$ to $N$}
                \If{\texttt{filtered\_masks}[j] is None} \textbf{continue} \EndIf
                \State Compute overlap $\gets$ \texttt{masks}[i] $\cap$ \texttt{masks}[j]
                \If{overlap small} \textbf{continue} \EndIf                    \State Add overlap to \texttt{new\_masks}, remove from masks
                \State Update masks, mark small ones for removal
                \State \texttt{updated} $\gets$ \textbf{True}
            \EndFor
        \EndFor
        \State Remove marked masks, add \texttt{new\_masks}            
    \EndWhile
    \State \Return \texttt{filtered\_masks}
    \end{algorithmic}
\end{algorithm}

\subsection{Prompt Engineering for GPT}
\textbf{System prompts:}

\begin{center}
\parbox{0.95\textwidth}{
\textbf{\texttt{You are controlling a microscope to scan a specific area. I want to complete the scan by calling the following functions:\\
1. `move\_by(dx, dy)` 2. Auto exposure and autofocus function 3. Capture and save image function 4. Image processing function\\
~\\
Task requirements:\\
- Move step-by-step, scan the entire area, with the area size specified by user.\\
- After each movement, auto exposure, autofocus, analyze the image to generate labeled masks.\\
- Calculate the mask area ratio for the category specified by user.\\
- Find the position with the highest ratio, provide the coordinate.\\
- Move to this position, raise ‘manually switch the lens to 100x’ and wait 10s. \\
- Call the auto exposure, autofocus, image analysis.\\
~\\
Once all steps are complete, the task is finished.}}
}
\end{center}

\textbf{Task prompts:}

\begin{center}
\parbox{0.95\textwidth}{
\textbf{\texttt{The current task is to scan a 2x2 grid where each cell is 266x224 in size, total area size for the scan is 532x448. Starting from the previous position: x = x$_0$, ~ y = y$_0$. Please calculate the mask area ratio for non-substrate categories and find the region with the highest ratio.}}
}
\end{center}

\textbf{Classification prompts:}

\begin{center}
\parbox{0.95\textwidth}{
\textbf{\texttt{Analyze these distortion values from K-means clustering and determine the optimal number of clusters using the Elbow Method. \\
Raw distortion values: \textbraceleft distortions\textbraceright \\
Percentage changes between consecutive points: \\
\textbraceleft [f'\textbraceleft x:.2f\textbraceright\%' for x in pct\_changes]\textbraceright \\
Please consider: \\
1. The point where adding more clusters gives diminishing returns (the "Elbow" point) \\
2. The percentage change in distortion between consecutive points \\
3. Balance between model complexity and explanation power \\
4. Practical considerations (avoid over-segmentation) \\
Rules: \\
- The optimal number must be between 2 and \textbraceleft len(distortions)\textbraceright \\
- Look for significant drops in improvement (usually less than 10-15\% improvement) \\
- Consider if the gain in adding another cluster justifies the increased complexity \\
- Factor in that this is for image segmentation, where too many segments can be impractical \\
Please respond with only the number representing the optimal cluster count. No explanation needed.}}
}
\end{center}


\newpage

\section{Supplementary Figures}
\begin{figure}[H]
    \centering
    \includegraphics[width=1\textwidth]{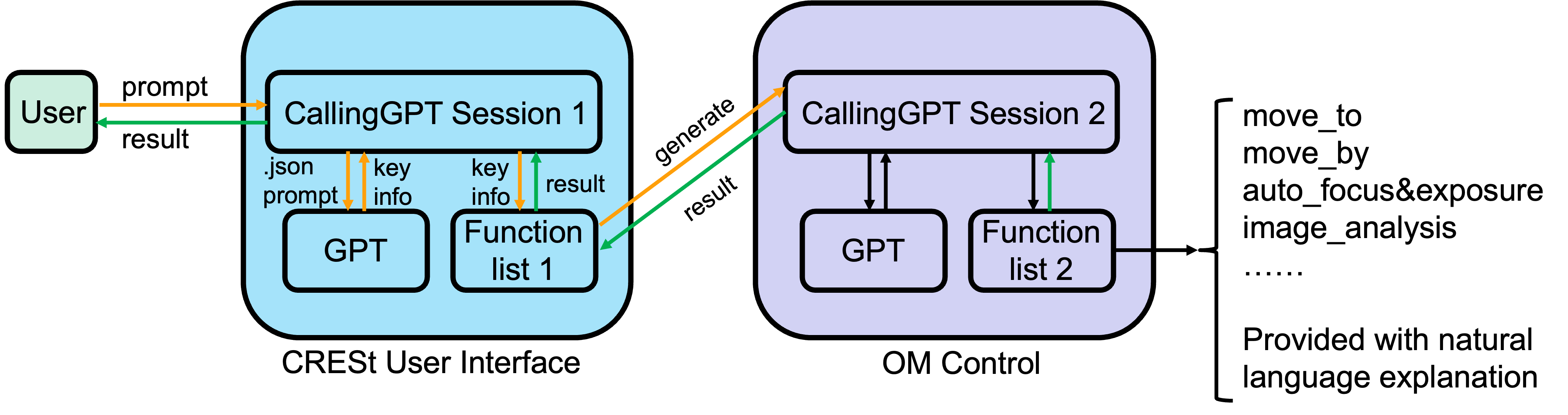}
    \caption{Workflow of microscope control enabled by GPT}
    \label{fig:callinggpt}
\end{figure}

\begin{figure}[H]
    \centering
        \centering
        \includegraphics[width=0.5\textwidth]{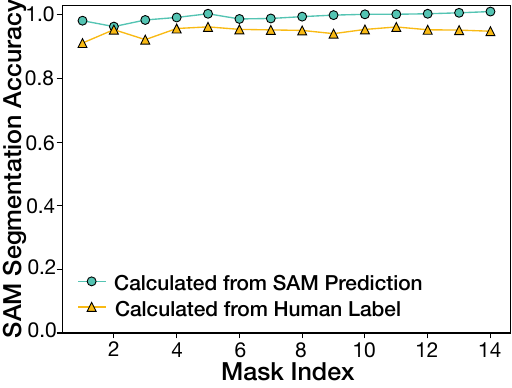}
        \caption{Comparison of SAM segmentation accuracy based on human labels and SAM predictions (Predict IoU) across multiple masks.}
        \label{fig:IoU}
\end{figure}

\begin{figure}[H]
    \centering
    \includegraphics[width=0.6\textwidth]{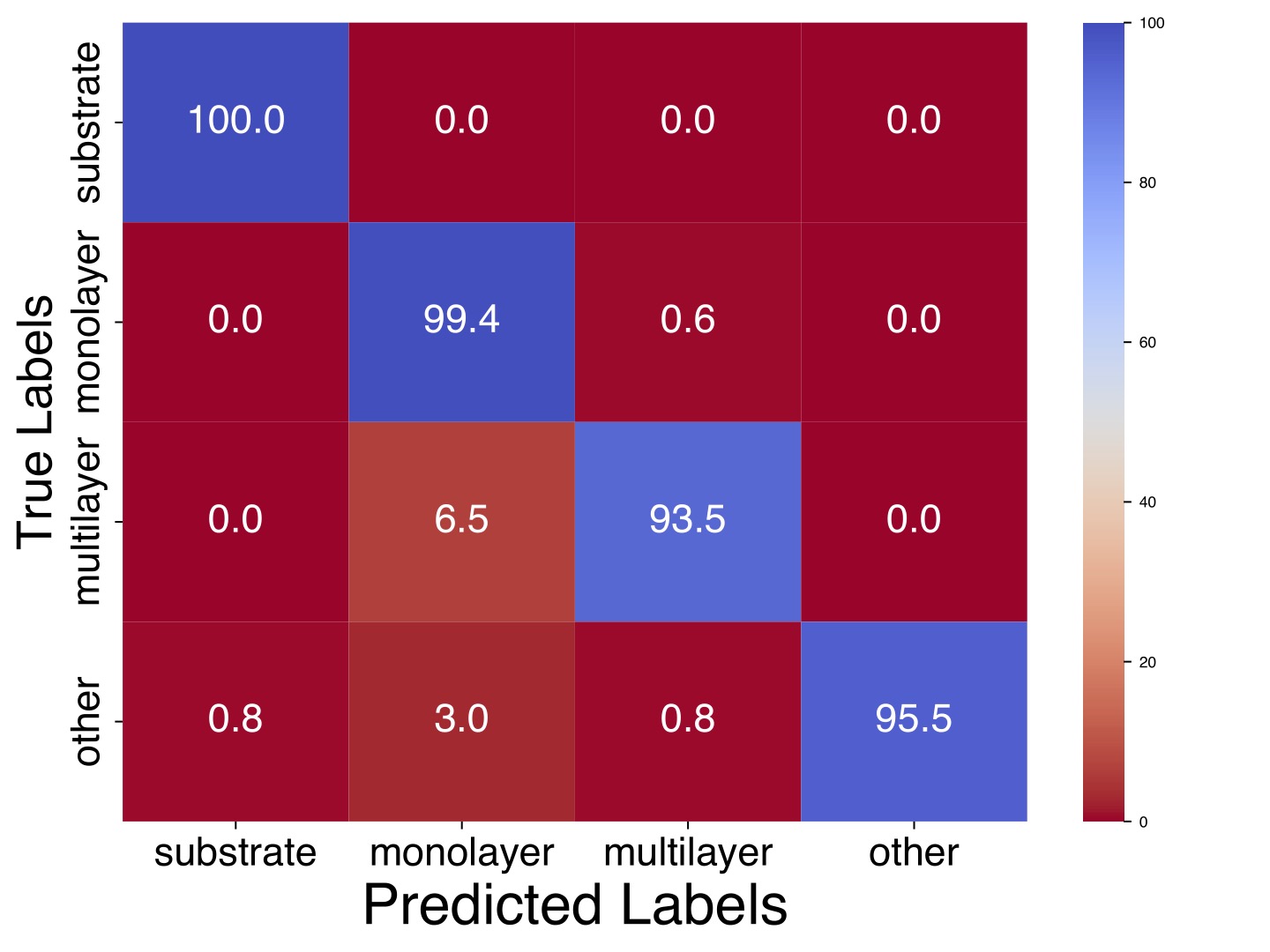}
    \caption{Mask-level confusion matrix of ATOMIC model}
    \label{figS3}
\end{figure}

\begin{figure}[H]
    \centering
    \includegraphics[width=1\textwidth]{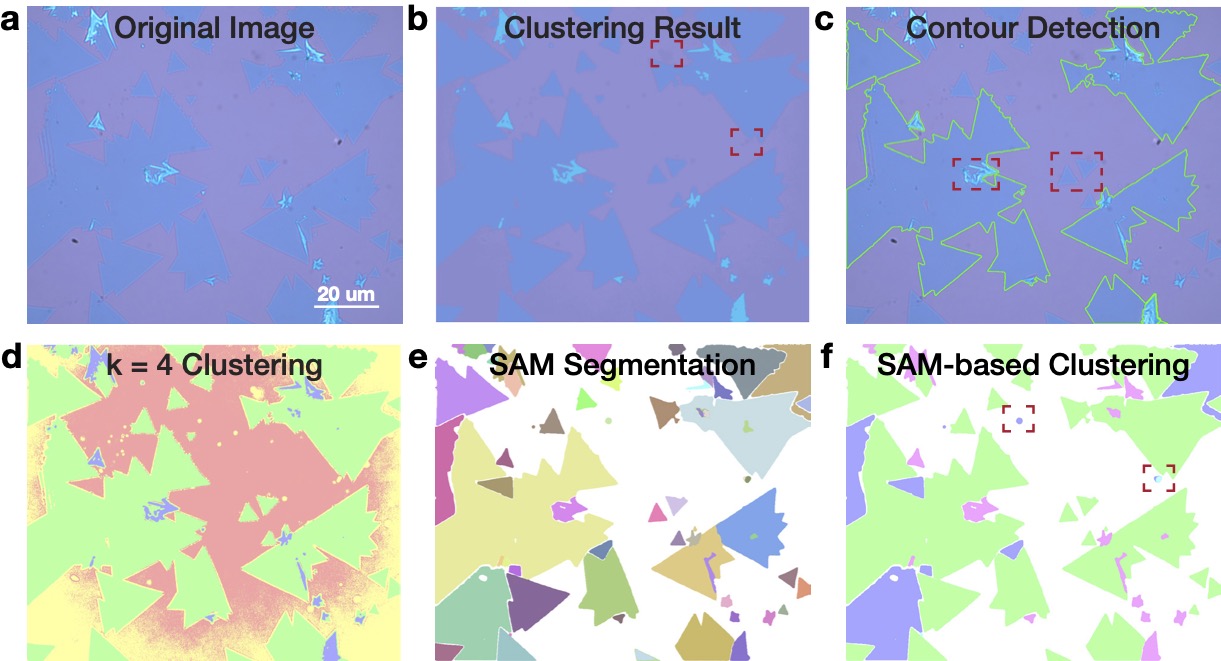}
    \caption{Performance comparison between our SAM + clustering approach and direct clustering + contour detection approach, highlighting the critical advantages of incorporating SAM in our autonomous synergy. a, Original microscopic image of CVD-grown MoS$_2$. b, Direct clustering result, with red boxes highlighting incorrect and undesired clusters. Each cluster is filled with the RGB color of its corresponding cluster center. c, Cluster-based contour detection, where red boxes indicate missed edge detections. This result is derived from b. d, Direct clustering result with human picked k=4. e-f, SAM segmentation and SAM-based clustering result, with red boxes highlighting its superior boundary preservation and segmentation accuracy.}
    \label{figS4}
\end{figure}

\begin{figure}[H]
    \centering
    \includegraphics[width=1\textwidth]{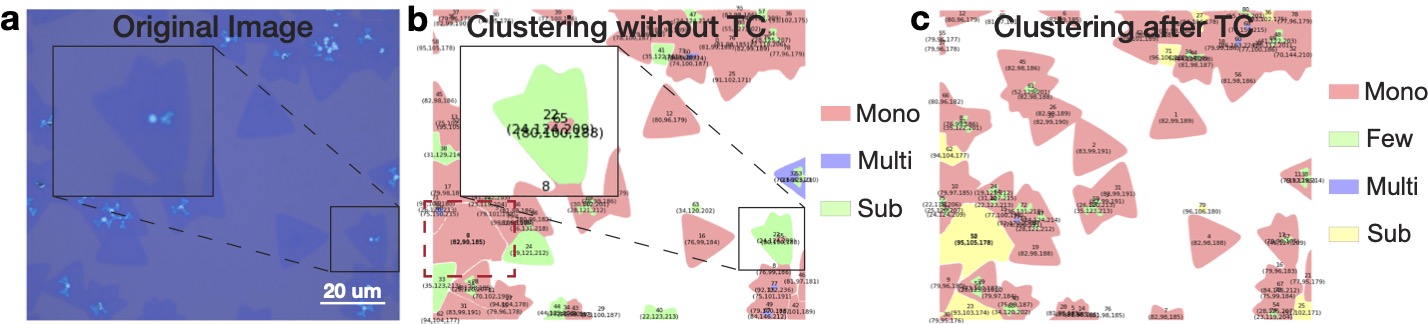}
    \caption{The role of Topological correction (TC) in clustering process. a, Original image of the sample, containing substrate, monolayer, few-layer, and multilayer regions. b, Clustering results without TC, where confusion occurs between the substrate and monolayer (red dashed box), and fine structures are not well distinguished (black solid box). c, Clustering results after applying TC, showing improved classification with the introduction of an additional few-layer (green) category and better separation of monolayer (red), multilayer (blue), and substrate (yellow) regions. The refined clustering enhances the accuracy of feature segmentation.}
    \label{figS5}
\end{figure}

\begin{figure}[H]
    \centering
    \includegraphics[width=1\textwidth]{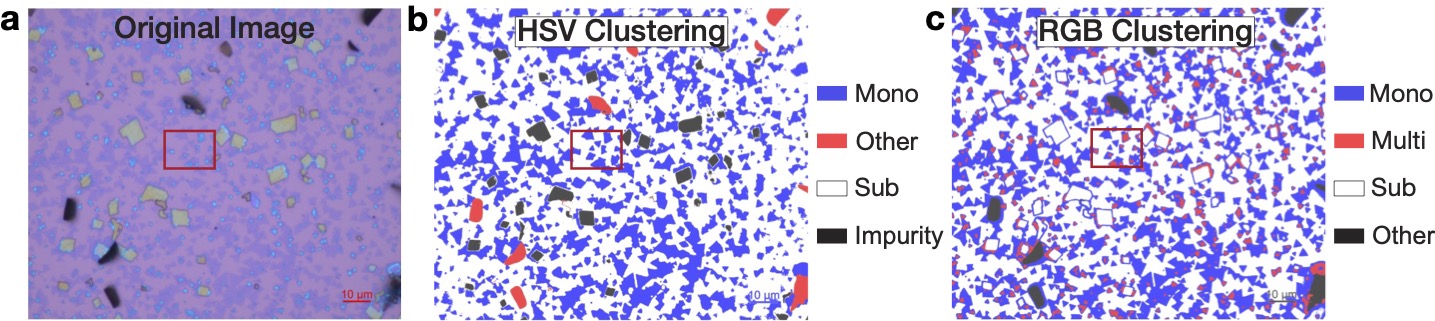}
    \caption{Comparison of Clustering results in RGB and HSV color spaces. a, Original image of the sample. b, HSV-based clustering result. c, RGB-based clustering result. The classified categories: monolayer (Mono), multilayer/crystal (Multi), substrate (Sub), impurities and others.}
    \label{figS6}
\end{figure}

\begin{figure}[H]
    \centering
    \includegraphics[width=\textwidth]{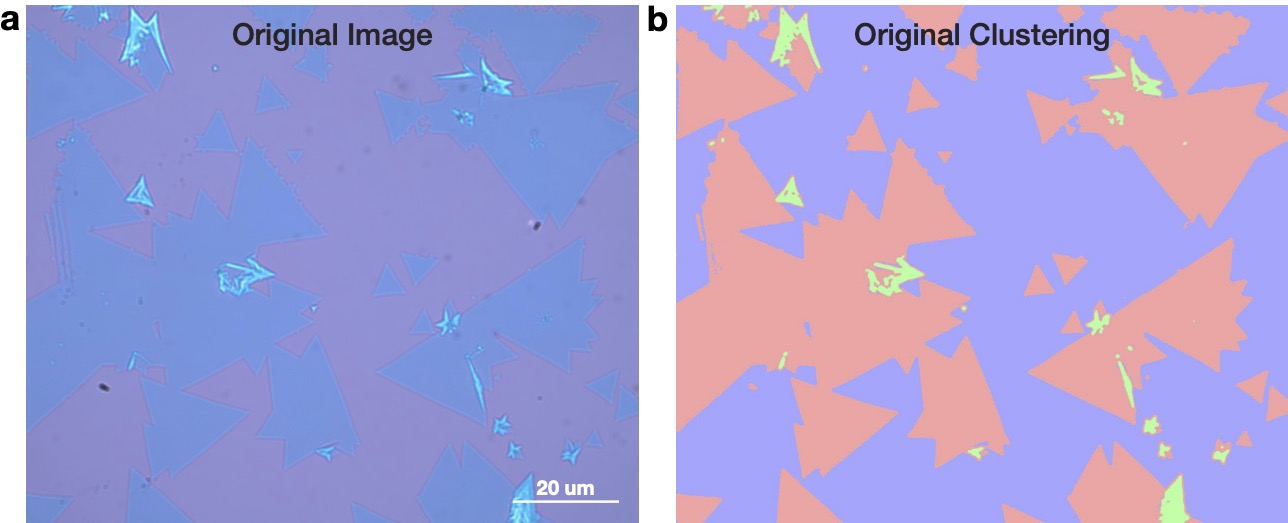}
    \caption{a, Original image of the sample. b, RGB-based pixel-wise clustering result. }
    \label{figS7}
\end{figure}

\begin{figure}[H]
    \centering
    \includegraphics[width=\textwidth]{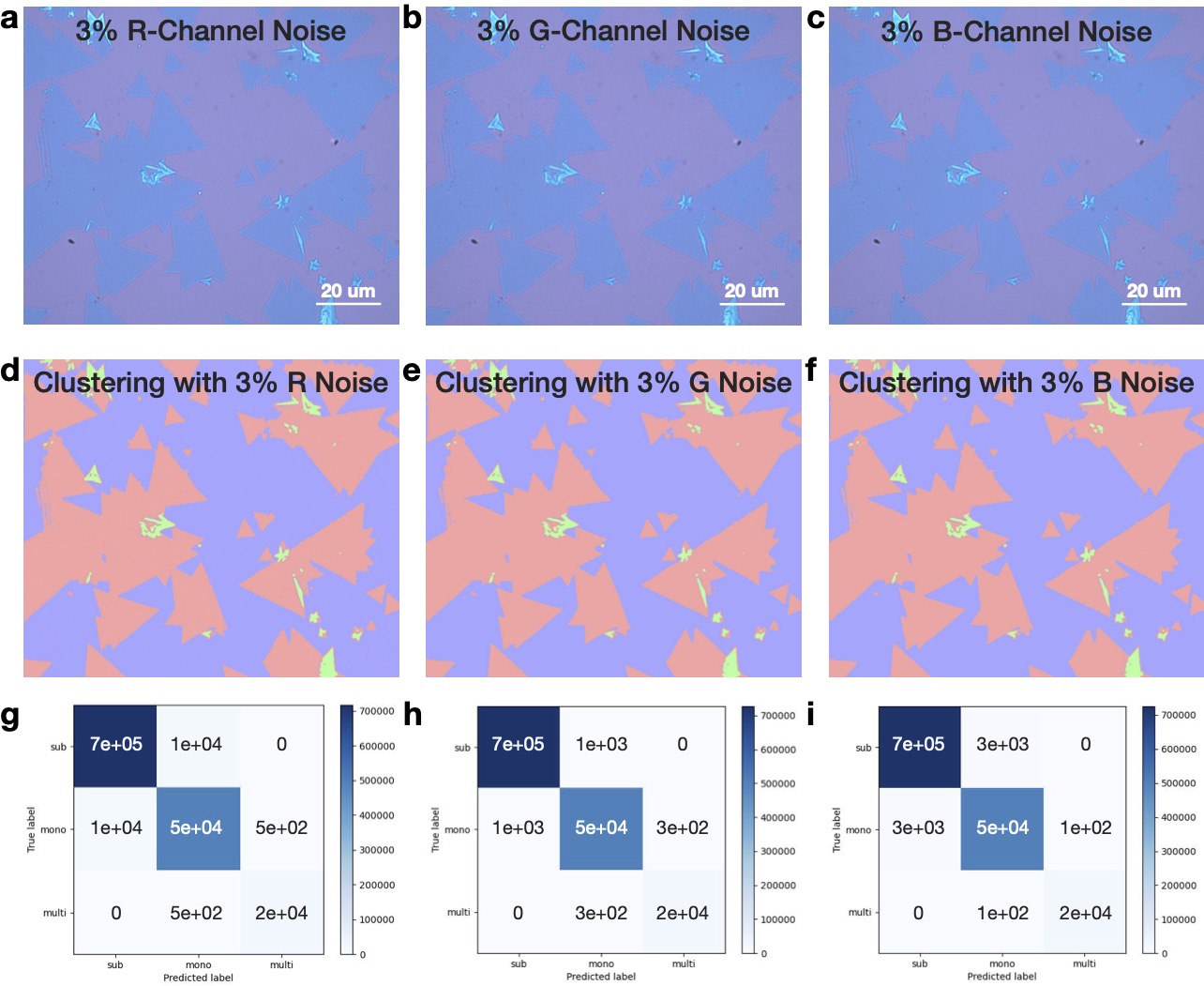}
    \caption{Clustering results after applying 3\% Gaussian noise to the R, G, and B channels of the original image. a-c show the noisy images for each channel, d-f display the corresponding clustering outcomes. g-i exhibit the confusion matrices calculated using the clustering result of the noise-free image in Fig.S7b as the ground truth.}
    \label{figS8}
\end{figure}

\begin{figure}[H]
    \centering
    \includegraphics[width=\textwidth]{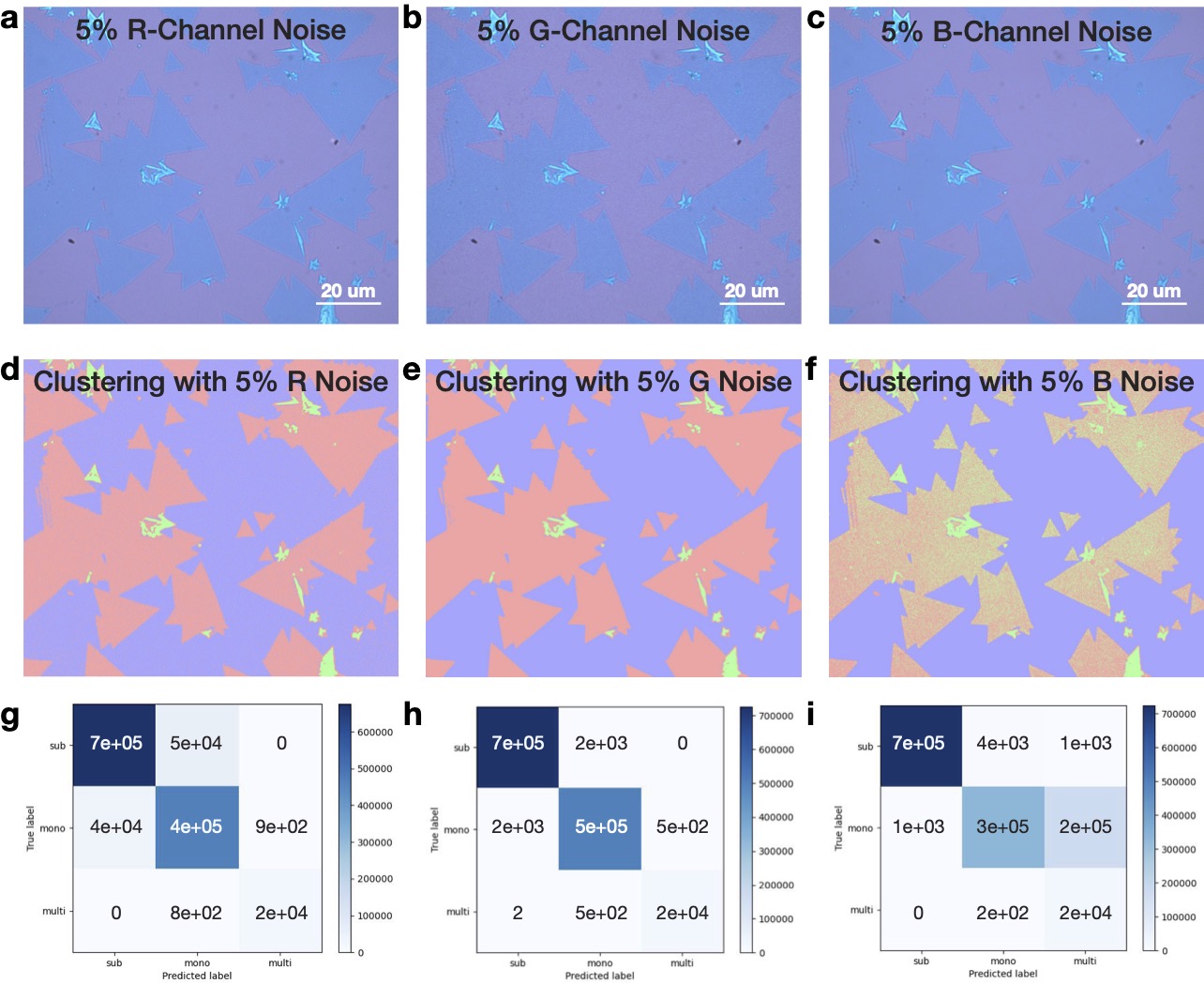}
    \caption{Clustering results after applying 5\% Gaussian noise to the R, G, and B channels of the original image. a-c show the noisy images for each channel, d-f display the corresponding clustering outcomes. g-i exhibit the confusion matrices calculated using the clustering result of the noise-free image in Fig.S7b as the ground truth.}
    \label{figS9}
\end{figure}

\begin{figure}[H]
    \centering
    \includegraphics[width=\textwidth]{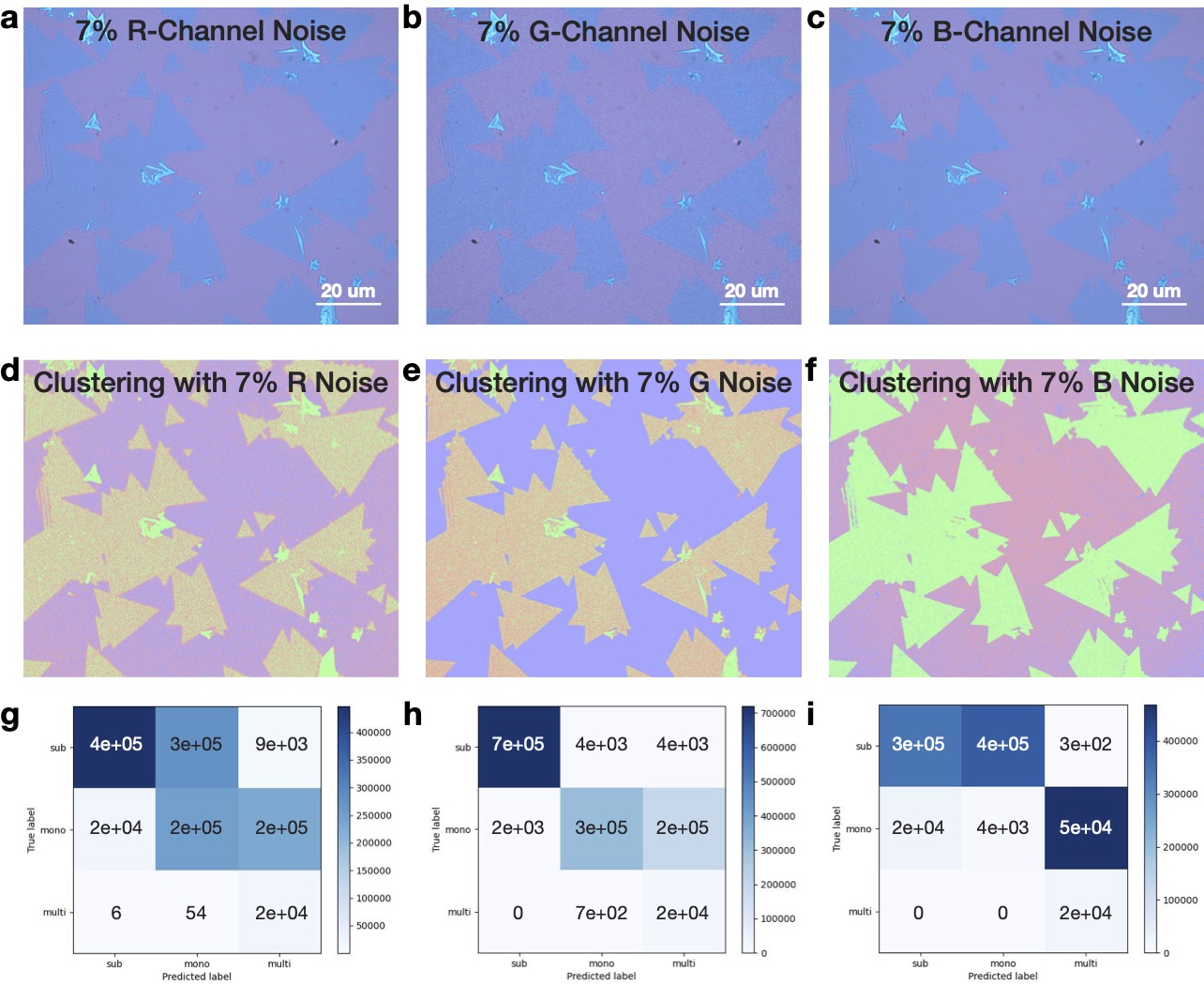}
    \caption{Clustering results after applying 7\% Gaussian noise to the R, G, and B channels of the original image. a-c show the noisy images for each channel, d-f display the corresponding clustering outcomes. g-i exhibit the confusion matrices calculated using the clustering result of the noise-free image in Fig.S7b as the ground truth.}
    \label{figS10}
\end{figure}

\begin{figure}[H]
    \centering
    \includegraphics[width=\textwidth]{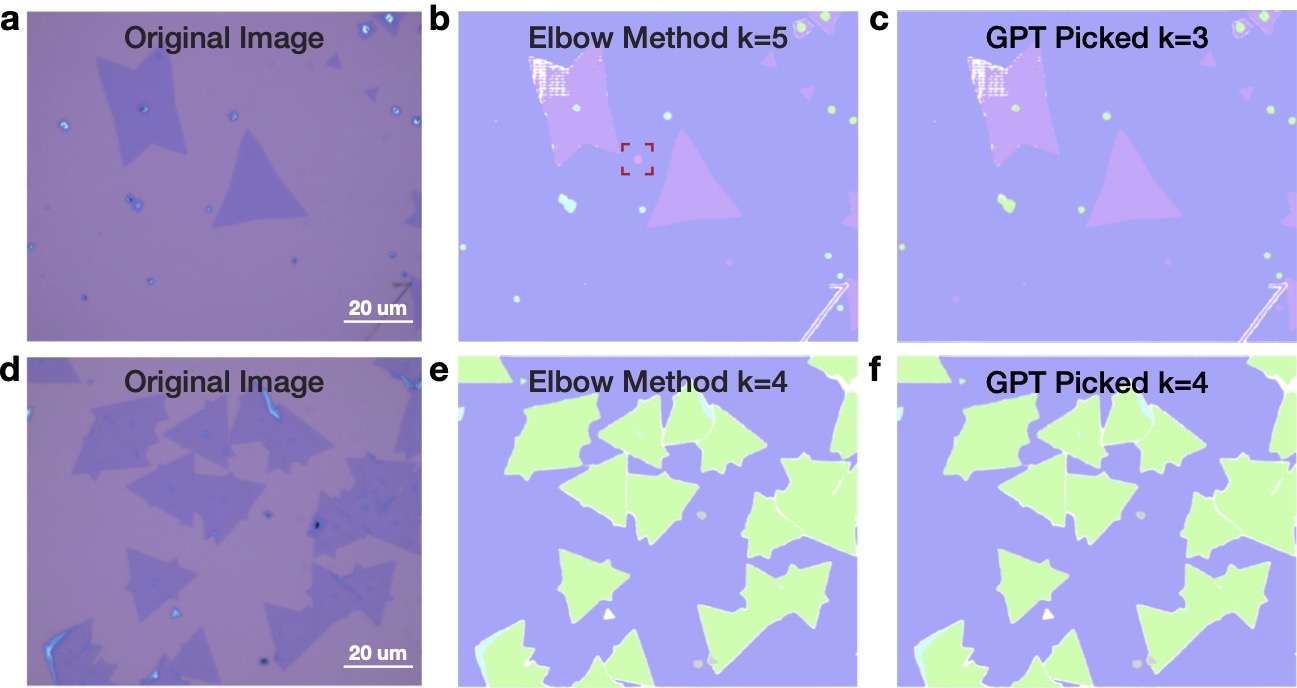}
    \caption{(a, d) Original om images of two samples. (b, e) Segmentation results using the Elbow method with k=5 and k=4, respectively. (c, f) Segmentation results using GPT-picked k values of 3 and 4, respectively.}
    \label{figS11}
\end{figure}

\begin{figure}[H]
    \centering
    \includegraphics[width=\textwidth]{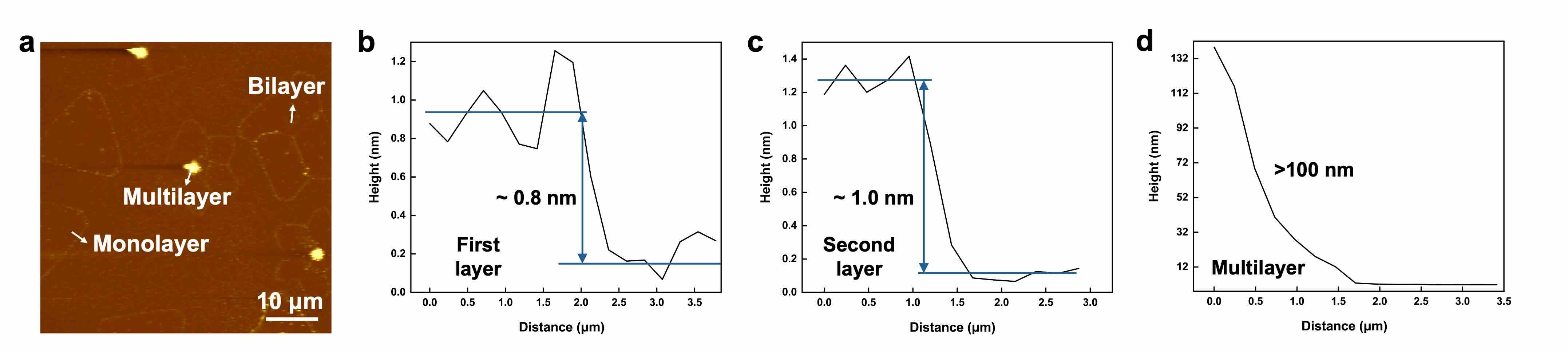}
    \caption{AFM image and corresponding height profiles of monolayer, bilayer, and multilayer MoS$_2$. a, the spatial distribution of the layers, labeled accordingly. b-d, the height profiles across selected regions, indicating thicknesses for the first layer and second layer corresponding to theoretical values, and above 100 nm thickness for the multilayer.}
    \label{figS12}
\end{figure}

\begin{figure}[H]
    \centering
    \includegraphics[width=0.75\textwidth]{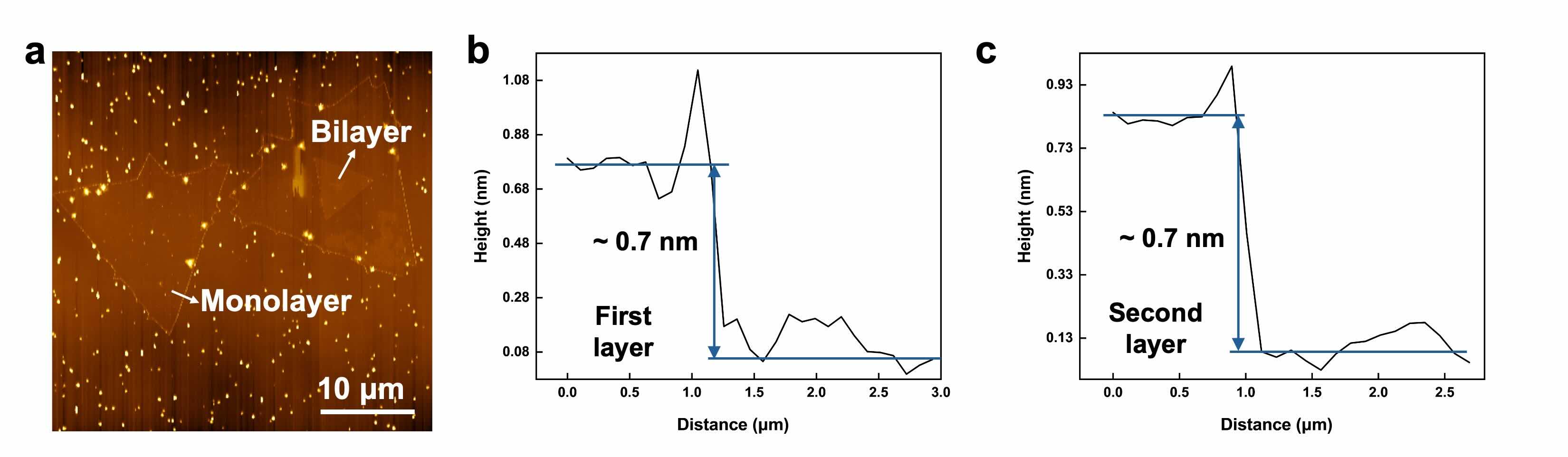}
    \caption{AFM image and corresponding height profiles of monolayer and bilayer WSe$_2$. a, the spatial distribution of the layers, labeled accordingly. b-c, the height profiles across selected regions, indicating approximate thicknesses of 0.7 nm for the first and second layers.}
    \label{figS13}
\end{figure}

\begin{figure}[H]
    \centering
        \centering
        \includegraphics[width=0.8\textwidth]{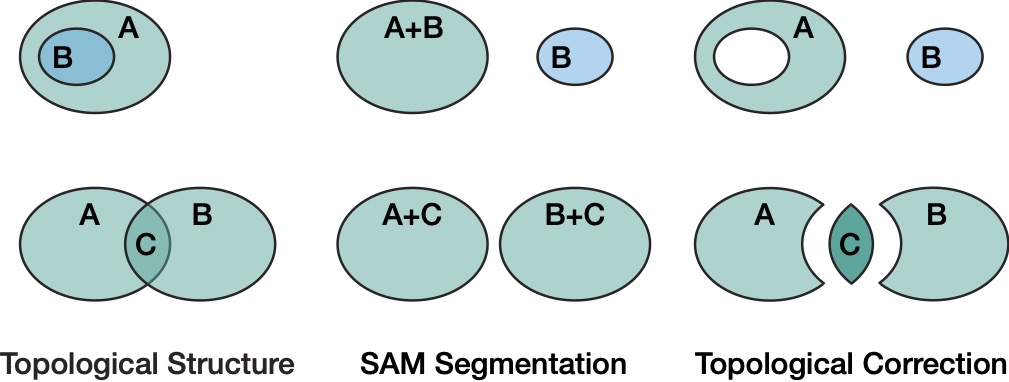}
        \caption{Topological correction in segmentation. The first column shows original structures with overlaps. The second column presents SAM segmentation results with merged regions. The third column illustrates corrected segmentation, restoring accurate boundaries and relationships.}
        \label{figS14}
\end{figure}

\newpage
\section{Supplementary Tables}
\begin{table}[h]
    \centering
    \caption{Mask-Level Confusion Matrix of ATOMIC model}
    \begin{tabular}{ccccc}
        \toprule
        True \textbackslash{} Prediction & Substrate & Monolayer & Multilayer & Other \\
        \midrule
        Substrate  & 147  & 0    & 0    & 0   \\
        Monolayer  & 0    & 683  & 4    & 0   \\
        Multilayer & 0    & 24   & 346  & 0   \\
        Other      & 1    & 4    & 1    & 127 \\
        \bottomrule
    \end{tabular}
\end{table}

\begin{table}[h]
    \centering
    \caption{Pixel-Wise Confusion Matrix of ATOMIC model}
    \begin{tabular}{ccccc}
        \toprule
        True \textbackslash{} Prediction & Substrate & Monolayer & Multilayer & Other \\
        \midrule
        Substrate  & 86,649,891 & 0          & 0         & 0         \\
        Monolayer  & 0          & 34,131,868 & 22,955    & 0         \\
        Multilayer & 0          & 310,422    & 3,893,999 & 0         \\
        Other      & 5,789      & 7,141      & 1,557     & 455,647   \\
        \bottomrule
    \end{tabular}
\end{table}